\documentstyle[epsfig]{mn}

\begin{document}
\def\ltsima{$\; \buildrel < \over \sim \;$}
\def\simlt{\lower.5ex\hbox{\ltsima}}
\def\gtsima{$\; \buildrel > \over \sim \;$}
\def\simgt{\lower.5ex\hbox{\gtsima}}

\title[Iron and Nickel properties in the X--ray reflecting region of the 
Circinus Galaxy]
{Iron and Nickel line properties in the X--ray reflecting region of the 
Circinus Galaxy}

\author[Silvano Molendi, Stefano Bianchi and Giorgio Matt]
{Silvano Molendi$^1$, Stefano Bianchi$^2$ and Giorgio Matt$^2$\\ ~ \\
$^1$IASF--CNR, Via Bassini 15, I--20133, Milano, Italy \\
$^2$Dipartimento di Fisica, Universit\'a degli Studi ``Roma 
Tre'', Via della Vasca Navale 84, I--00146 Roma, Italy \\
}

\maketitle
\begin{abstract}
We discuss the iron and nickel properties in the nuclear X-ray 
reflecting region of the Circinus Galaxy, studied with
XMM-$Newton$. The main results
are: a) from the depth of the Fe K$\alpha$ edge, a value of $A_{\rm Fe}$=1.7 in number 
with respect to the cosmic value (as for Anders \& Grevesse 1989) is measured, if
the (not directly visible) illuminating spectrum
is assumed to be that measured by
BeppoSAX. If the slope of the primary power law
is left free to vary, a steeper spectrum and a lower iron abundance
(about 1.2) are found. b) From the Ni to Fe  K$\alpha$
lines flux ratio, a nickel-to-iron abundance ratio of 0.055-0.075 is found. 
c) The Fe K$\beta$/K$\alpha$ flux ratio is  slightly lower than
expected, possibly due to a mild ionization of iron (which however cannot
be much more ionized than {\sc x}).
d) The presence of the Fe K$\alpha$  Compton Shoulder, already discovered
by $Chandra$, is confirmed,
its relative flux implying Compton--thick matter. This  
further supports the identification of the reflecting region with the absorber.

\end{abstract}


\section{Introduction}

The Circinus galaxy, at the distance of about 4 Mpc, 
is the closest and brightest among
Compton-thick Seyfert 2 galaxies (see Matt 2002a and Matt et al. 2000
for reviews).  Its
nuclear 2--10 keV X--ray spectrum is dominated by reflection from a spatially 
unresolved region (which hereinafter we will call the `torus',
Antonucci 1993) with
a size less than about 15 pc (Sambruna et al. 2001a). Most of the line
spectrum (Sambruna et al. 2001b; Guainazzi et al. 1999; Matt et
al. 1996) can be explained in terms of a single, mildly ionized 
reflector (Bianchi et al. 2001) even if a second, more ionized
reflecting region is required to account for the faint He--like Fe K$\alpha$
line detected by the $Chandra/HETG$ (Sambruna et al. 2001b). 
Given the brightness of the source and the dominance of the cold
reflection component, the Circinus Galaxy is ideal to study
the physical, chemical and geometrical properties of the torus.

Above 10 keV, the nuclear emission becomes visible, piercing through a cold
absorber with $N_{\rm H}=4\times10^{24}$ cm$^{-2}$ (Matt et al. 1999).
It is natural to identify the absorber with the reflector, as the
latter should be Compton--thick, as implied by its spectrum (e.g. Matt et al. 2003) 
and by the relative
flux of the Compton Shoulder (Bianchi et al. 2002; Matt 2002b; this
paper), i.e. the part of the line profile due
to photons which are scattered before escaping the matter.
 With this assumption, from the modeling of the line spectrum
a value of about 0.2 pc for the inner radius of the torus can be
derived (Bianchi et al. 2001).

In this Letter we analyze and discuss the hardest part of the
spectrum with the aim of studying the iron and nickel properties, 
taking advantage of the unprecedented sensitivity of  XMM--$Newton$
at these energies. A complete spectral and spatial analysis is beyond
the scope of this paper, and is deferred to a future work.

\section{Data reduction}

The Circinus Galaxy was observed by XMM--$Newton$ for about 110 ks on 
August 6 and 7 2001. The EPIC p-n and  MOS1 cameras were operated in  
full frame mode
while the EPIC MOS2 was operated  in  large window mode.  ODF files
were downloaded from the public archive and processed using version
5.4.1 of the SAS software.  Further screening for soft  proton  flares
and residual hot
pixels   was  performed  using    procedures  described in   Baldi  et
al. (2002). The  effective exposure times after cleaning are 75, 77 and 70
ks for MOS1, MOS2 and p-n respectively.
For our spectral analysis we  used pattern 0 to 12 events
for MOS1 and  MOS2 while for  p-n we accumulated separately single and
double events spectra.  
Spectra for all three cameras were extracted from a 30$^{\prime\prime}$ radius
circle centered on the source peak. 
Two  nearby sources lie within this radius, but were excluded  by
excising  circles  with a 7$^{\prime\prime}$   radius centered  on the
respective peaks.  Inspection  of a 6-10  keV image shows  that the
intensity at  the peak for these sources is less  then a 10-15\% of 
the Circinus peak intensity. Exclusion  of their  core is  therefore
sufficient to make contamination very small.  
Background  spectra for the  3 cameras were extracted  from annuli
with bounding radii 90$^{\prime\prime}$ and 150$^{\prime\prime}$
centered on the source, after having verified that in these regions
no detectable  sources were present.
Redistribution  matrices and  effective areas were generated
using the  {\sc rmfgen}  and {\sc arfgen} tools.  Above $\sim$2 keV the standard
energy binning for matrices generated by {\sc rmfgen} is 50 eV  for p-n and 5
eV for MOS. Because our aim  is to perform detailed spectroscopy on
Fe  and Ni  lines observed between  6  and 8 keV,  we  have chosen  a finer
binning  of 1 eV for both MOS  and p-n  matrices.  The effective areas
were of course generated with  the same binning. For the fitting procedure, we rebinned
the data to have at least 20 counts per bin, in order to ensure applicability of $\chi^2$
statistics.

\section{Data analysis and results}

\begin{figure}
\epsfig{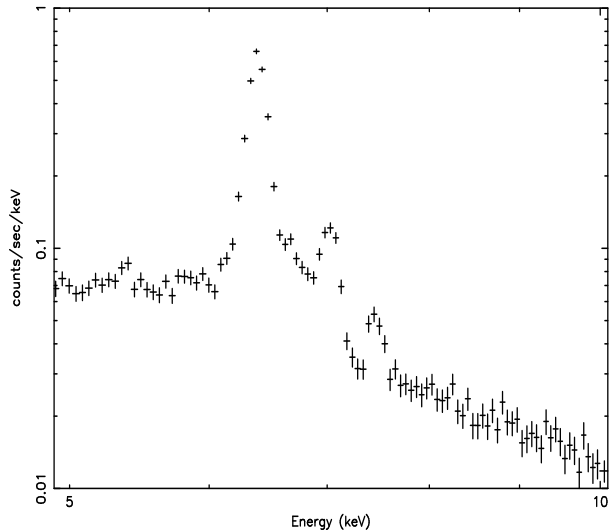}
\caption{The 5-10 keV p-n spectrum.}
\label{spectrum}
\end{figure}

In Fig.~\ref{spectrum} the (p-n, single event) 
spectrum of the nucleus above 5 keV is
shown. Three emission lines are obviously present,
readily identified with iron K$\alpha$ and K$\beta$ 
as well as nickel K$\alpha$. The expected energies of these lines,
{\sl assuming neutral atoms}, are:
6.400 keV (actually a doublet, composed by a  
6.391 keV and a 6.405 keV line);
7.058 keV (a doublet, with energies of 7.057 keV and 7.058 keV)
 and 7.472 keV (a doublet, with energies of 7.461 keV and 7.478
keV). All line energies are from Kaastra \& Mewe (1993).
For each doublet, the intensity ratio between the two lines is 1:2,
and the given energies are the weighted mean.

In the following, unless explicitely stated, we will adopt the cosmic element
abundances given by Anders \& Grevesse (1989), which are 
4.68$\times$10$^{-5}$ for iron and 1.78$\times$10$^{-6}$ for nickel
(by number with respect to hydrogen); and the fluorescent yields 
given by Kaastra \& Mewe (1993), which for neutral atoms are:
0.304 (iron K$\alpha$); 0.038 (iron K$\beta$); 0.368 (nickel K$\alpha$).

In the following, 
we will also assume that the absorbing material is dust-free. In fact, from the 
observed N$_{\rm H}$ of 4$\times10^{24}$ cm$^{-2}$ a value of $A_{\rm V}$=2000 is
expected for a gas-to-dust ratio like that in the ISM of our own galaxy (Savage
\& Mathis 1979), while
the derived value from IR observations is 50 only (Moorwood 1996).

We fitted separately the spectra from MOS1, MOS2, p-n single events
and double events. The p-n single events  gives 
values of the line energies close to the expected ones for neutral
atoms (the other
instruments giving values systematically higher, MOS1 being the
most discrepant, giving e.g. 6.436 for the iron K$\alpha$). Therefore,
in the following we will limit the analysis to this
instrument which is the one with the larger
signal--to--noise. In principle, as it is 
possible that the matter is mildly ionized, this choice is somewhat 
arbitrary, but it is supported by the K$\beta$ line flux, as explained
in Sec. 3.3.

\begin{figure}
\epsfig{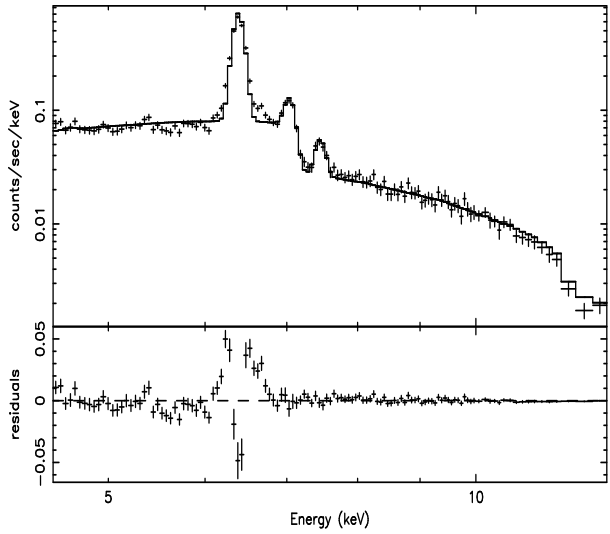}
\caption{The best fit spectrum and residuals when fitted with
a cold reflection component plus three narrow gaussian lines accounting
for the iron K$\alpha$ and K$\beta$ lines, and the nickel K$\alpha$
line.}
\label{wings}
\end{figure}

\begin{figure}
\epsfig{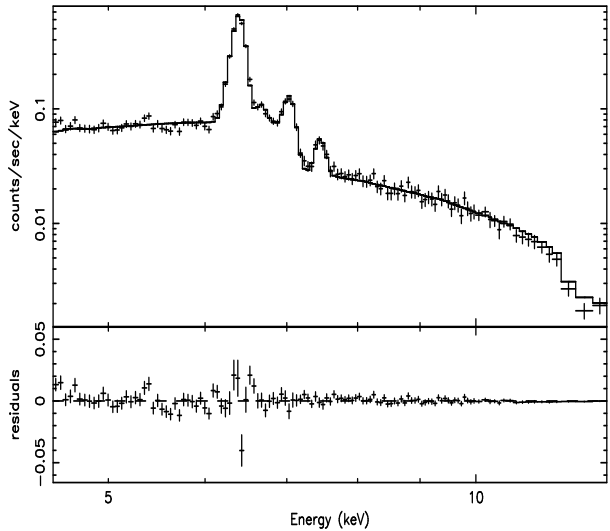}
\caption{The best fit spectrum and residuals when the Compton
Shoulder and the He--like iron K$\alpha$ line are included.}
\label{good_sp}
\end{figure}

The spectrum in the 4.5 to 13 keV range has been fitted with a model
composed of a cold reflection component (model {\sc pexrav} in
XSPEC), produced by a power-law with $\Gamma=1.56$ and a exponential
cutoff at $E_{\mathrm{C}}=56$ keV (Matt et al. 1999, Bianchi et al. 2001),
and the three emission lines  mentioned above, all of them
described by a gaussian model with $\sigma$=1 eV (i.e. much less than
the instrument energy resolution) and a redshift of 
0.0015 (Freeman et al. 1997). The $\chi^2$ is poor (381/117 d.o.f.), largely
due to residuals around the iron K$\alpha$ line (see
Fig.~\ref{wings}). Leaving the width of the line free to vary, the
fit improves significantly ($\chi^2$=206/116 d.o.f.), but residuals
are still visible, and the width of the line, $\sigma=42$ eV, is not
easy to explain if we assume that the line emitting matter is 
the molecular torus\footnote{If the line is emitted by a torus rotating
around the black hole with
keplerian velocity, and assuming a black hole mass of 1.7$\times10^6 M_{\odot}$ (Greenhill
et al. 2003) and an inner radius of the torus of 0.2 pc (Bianchi et al. 2001),
the expected value of $\sigma$ is about 2 eV.}. It seems more likely that the line
`wings' are instead due to the Compton 
shoulder (Bianchi et al. 2002), redwards of the line core, and 
to the He--like iron line (Sambruna et al. 2001b), bluewards of the
line core. The Compton shoulder is indeed expected on theoretical ground
(Matt 2002b), while the  He--like iron line probably originates from
reflecting circumnuclear ionized matter.
We therefore added two more lines, with energies fixed
at 6.3 keV (Matt 2002b) and 6.68 keV, respectively (the width of the
line core, as well as that of the added lines, held fixed to 1 eV
for simplicity). The fit
improves significantly ($\chi^2$=157/115 d.o.f.), and no strong
systematic residuals are left (Fig.~\ref{good_sp}). The significance
of each line is $>$99.99\%, according to the F-test. The Compton shoulder
is actually expected to have a finite width, but unfortunately
its profile (Sunyaev \& Churazov 1996, Matt 2002) 
is different from any {\sc xspec} model. As a check, we tried for the Compton
Shoulder also gaussians with $\sigma$=40 and 70 eV (the latter value corresponding
to a FWHM of the same order of the total width of the Shoulder). A 
worse fit ($\chi^2$ of 162/115 and 179/115, respectively) is obtained, 
but the fluxes of both the Shoulder and the line core remain almost unchanged.

The best fit results are summarized in Table~\ref{bestfit}.
All errors refer to 90\% confidence level for one interesting
parameter.  It is worth noting that the measured iron K$\alpha$ line energy
constrains, at the above confidence level, the ionization state
of iron (before the photoionization) to be between {\sc xii} and {\sc xiii}
(House 1969), or to be exactly {\sc xvii} if the MOS1 value is
considered. While of course, given the present uncertainties on the
instruments calibrations, these results cannot be taken too literally,
they highlight the potentially extraordinary precision of the measure.
More solid constraints on the ionization states can be derived from the
Fe K$\beta$/K$\alpha$ lines ratio, as discussed in Sec.~3.3.

\begin{table}[t]
\caption{Best fit results} 
\begin{tabular}{||l|c|||}
\hline
& \cr
$\Gamma$ & 1.56 (fixed) \cr
& \cr
$E_{\rm c}$ [keV] & 56 (fixed) \cr
& \cr
A$_{\rm Fe}^{a}$ & 1.72$^{+0.14}_{-0.09}$ \cr
& \cr
E (Fe K$\alpha$) [keV] & 6.413$^{+0.002}_{-0.002}$ \cr
& \cr
F (Fe K$\alpha$) [10$^{-5}$ ph cm$^{-2}$ s$^{-1}$] & 21.4$^{+0.6}_{-0.6}$ \cr
& \cr
EW (Fe K$\alpha$) [eV] & 1160 \cr
& \cr
E (Fe K$\beta$) [keV] & 7.051$^{+0.008}_{-0.009}$ \cr
& \cr
F (Fe K$\beta$) [10$^{-5}$ ph cm$^{-2}$ s$^{-1}$] & 2.94$^{+0.23}_{-0.24}$ \cr
& \cr
EW (FeK $\beta$) [eV] & 201 \cr
& \cr
E (Ni K$\alpha$) [keV] & 7.466$^{+0.013}_{-0.014}$ \cr
& \cr
F (Ni K$\alpha$) [10$^{-5}$ ph cm$^{-2}$ s$^{-1}$] & 1.49$^{+0.18}_{-0.24}$ \cr
& \cr
EW (Ni K$\alpha$) [eV] & 175 \cr
& \cr
F (Fe He-like K$\alpha$) [10$^{-5}$ ph cm$^{-2}$ s$^{-1}$] & 
1.36$^{+0.25}_{-0.24}$ \cr
& \cr
EW (Fe He--like K$\alpha$) [eV] & 72 \cr
& \cr
F (Fe K$\alpha$ CS) [10$^{-5}$ ph cm$^{-2}$ s$^{-1}$] & 4.28$^{+0.41}_{-0.55}$ \cr
& \cr
\hline
\end{tabular}
~\par
$^{a}$ in solar units (Anders \& Grevesse 1989) by number.
\label{bestfit}
\end{table}

\subsection{The Fe abundance}

The iron abundance is directly measured by the depth of the iron
edge in the Compton reflection continuum, and it is measured with
respect to the elements responsible for the photoabsorption below the 
edge, mainly oxygen and neon but with significant contributions from
magnesium, silicon and sulphur. The high quality of the
spectrum permits to estimate this parameter with high (statistical)
precision, i.e. less than 10\%. The result, i.e $A_{\rm Fe}\sim$1.7, 
has been obtained assuming an illuminating spectrum
as that derived from the BeppoSAX observation (Matt et al. 1999).
Leaving the photon index free to vary (and with the high energy
exponential cut--off fixed at 100 keV), a significantly better fit
is found ($\chi^2$=138/114 d.o.f., the improvement being significant at
the 99.98\% confidence level), and the iron abundance decreases to
about 1.2, a value which is more consistent with the iron
K$\alpha$ EW (e.g. Matt et al. 2003 and references therein). The best
fit power law photon index is 1.90$\pm$0.05.

In all fits described so far the
inclination angle of the reflector has been held fixed to the XSPEC
default value of 63$^{\circ}$. Leaving this parameter free to vary no
significant difference in the value of $A_{\rm Fe}$ is found.

In conclusion, given the uncertainties on the primary continuum as
well as on the geometry of the reflector (the {\sc pexrav} model is
for a plane--parallel slab), we can conclude that the iron abundance is
confined to be between 1 and 2. This is in agreement with the
results obtained by comparing the line EW and the Compton reflection
continuum in a sample of Seyfert 1 galaxies (Perola et al. 2002).

\subsection{The Ni abundance}

In order to derive, from the observed Ni to Fe K$\alpha$ line fluxes, the 
relative abundances of the two elements, we calculated the expected
values of their ratio, using the method 
proposed by Basko (1978) and valid for a semi--infinite
plane--parallel slab that is isotropically illuminated. For the nickel
line we included in the calculations both the unscattered and once
scattered line photons, while for the iron line, for which the two
components are separated in the fit procedure, 
we included only photons in the line core. The values of the 
calculated ratio range from 0.03 to 0.045, depending on the
inclination angle and the assumed power law index. The measured
value, 0.07 with an error of less than 20\%, is significantly larger,
strongly indicating a nickel-to-iron overabundance. In the
calculations, we adopted cosmic abundances. However, the results
do not depend much on the iron and nickel abundances, provided 
that they vary together. In fact, if e.g. both iron and
nickel are twice the cosmic value, we calculate that the flux of the nickel 
line increases only by a factor $\sim$1.3 (because of the increased
photoelectric absorption), similar to the increase
expected for the flux of the iron line (Matt et al. 1997), 
so leaving the need for a nickel-to-iron overabundance unchanged.

It must also be noted that, as we fitted the Ni line with a narrow
gaussian, part of its Compton Shoulder could be lost in the fitting
procedure, so possibly
further increasing the relative Ni overabundance.

Finally, we recall that in the calculations we have used the 
Anders \& Grevesse (1989) abundances. If we instead use the 
Anders \& Ebihara (1982) set, which differs mainly in having the
iron abundance 1.4 times lower, the expected values are of course
not much different from what we measure. The same is true if we use
the Grevesse \& Sauval (1998) or, even more, the Holweger (2001) sets. 
Put it in a different way,
our results implies that the nickel-to-iron absolute abundance ratio
is around 0.055-0.075, independently of course of the set of solar
abundances chosen for reference.

\subsection{The Fe K$\beta$/K$\alpha$ ratio}

The ratio between the Fe K$\beta$ and K$\alpha$ (core only) lines is 0.14
with a statistical error of about 10\%. This number slightly increases,
but within the statistical error, 
when the power law index of the illuminating continuum is kept free.
The expected value, again using the Basko (1978) formulae, is
0.155-0.16 (depending on the inclination angle), only marginally
consistent with the measured value. Again, it is possible that in the fit
procedure part at least of the Fe  K$\beta$ Compton Shoulder is lost,
so reducing the discrepancy. Alternatively, it is possible that the
actual value of the K$\beta$/K$\alpha$ probabilities is slightly lower
(for instance, Kikoin (1976) quoted a ratio of 17:150), or that the
iron is moderately ionized; interestingly, the observed value is
the one expected from Fe {\sc ix} or {\sc x} (Kaastra \& Mewe 1993), 
close to what  found from the p-n single
event iron K$\alpha$ line energy. Instead, a too small value is
expected for more ionized atoms, and of course no K$\beta$ line at
all is expected for Fe {\sc xvii}  (the ionization state found
from the MOS1) or more, when M electrons are no longer present.

\subsection{The Compton Shoulder}

The ratio between the iron  K$\alpha$  Compton Shoulder  and the line
core is 20$\pm$3\%, fully consistent with the value found by
$Chandra$ (Bianchi et al. 2002). The greater precision of the present
estimate with respect to the $Chandra$ one strengthens the case for 
a Compton--thick reflector (Matt 2002b), giving further support to 
the identification of the reflector with the absorber. It is important
to note that, while absorption provides information on the optical depth
of the line-of-sight material, features in reflection, being integrated over
the whole visible part of the matter, give information on the average value of the 
optical depth. The very fact that the amount of Compton shoulder is consistent
with the optical depth derived from absorption suggests that the material is
fairly homogeneous, so supporting the classical toroidal model for the cold
circumnuclear matter\footnote{See Matt \& Guainazzi 2003 for 
a case in which the optical depth derived from the Compton Shoulder is an order
of magnitude lower than that of the intervening matter.}.

\section{Summary}

The results discussed in this paper may be summarized as follows:

a) From the Fe K$\alpha$ edge a value of $A_{\rm Fe}$=1.7 in number 
with respect to solar
(using the set of values given by Anders \& Grevesse 1989) has been
found, assuming the primary illuminating continuum measured by
BeppoSAX (Matt et al. 1999). If the slope of the primary power law
is left free to vary, a steeper spectrum and a lower iron abundance
(about 1.2) is found.

b) From the flux ratio of the Ni to Fe  K$\alpha$ lines, a
nickel-to-iron abundance ratio of 0.055-0.075 is found, i.e. 
a factor of 1.5-2 Ni relative overabundance compared 
to the Anders \& Grevesse (1989) cosmic values, and instead
roughly consistent with the Anders \& Ebihara (1982) set. 
To the best of our knowledge this is the first unambiguous X--ray detection 
of a Ni line in an AGN. Previous detections, like in the ASCA spectrum of
Ark 564 (Turner et al. 2001) and in the BeppoSAX spectrum of Circinus itself
(Guainazzi et al. 1999) were affected by confusion with the Fe K$\beta$ line.

c) The Fe K$\beta$/K$\alpha$ flux ratio is $\sim$0.14, somewhat lower than
expected. A possible explanation is that the iron is mildly ionized,
but not much more than Fe {\sc x}, otherwise the Fe K$\beta$ line emission 
would be much fainter then observed (and of course absent at all for Fe {\sc xvii} or more).

d) The presence of the Fe K$\alpha$  Compton Shoulder, 
already discovered by Bianchi
et al. (2002) in the $Chandra$/HETG spectrum, is confirmed. The
relative flux of the Compton Shoulder is 20$\pm$3\%, fully consistent
with the $Chandra$ finding, and implying Compton--thick matter (Matt
2002b). This provides further support to the identification of the 
reflecting region with the absorber and suggests that the 
torus is fairly homogeneous.

\section*{Acknowledgments}

GM and SB acknowledges ASI and MIUR (under
grant {\sc cofin-00-02-36}) for financial support.

{}

\end{document}